\documentstyle[preprint,aps,prd]{revtex}

\begin{document}

\title {Heat Kernel and Scaling of Gravitational Constants}

\author{Diego A.\ R.\ Dalvit}

\address{{\it
Departamento de F\'\i sica, Facultad de Ciencias Exactas y Naturales\\
Universidad de Buenos Aires- Ciudad Universitaria, Pabell\' on I\\
1428 Buenos Aires, Argentina}}

\author{Francisco D.\ Mazzitelli}

\address{{\it
Departamento de F\'\i sica, Facultad de Ciencias Exactas y Naturales\\
Universidad de Buenos Aires- Ciudad Universitaria, Pabell\' on I\\
1428 Buenos Aires, Argentina\\
and\\
Instituto de Astronom\'\i a y F\'\i sica del Espacio\\
Casilla de Correo 67 - Sucursal 28\\
1428 Buenos Aires, Argentina}}

\date {February 1995 - Revised version}

\maketitle

\begin{abstract}
We consider the non-local energy-momentum tensor of quantum scalar
and spinor fields in $2 w$-dimensional curved spaces. Working to lowest order
in the curvature we
show that, while the non-local terms proportional to $\Box {\cal R}$,
$\Box \Box{\cal R}$, $\ldots, \Box^{w-2} {\cal R}$ are fully
determined by the early-time
behaviour of the heat kernel, the terms proportional to ${\cal R}$ depend on
the asymptotic late-time behaviour.
This fact explains a discrepancy between the running of the Newton constant
dictated by the RG equations and the quantum corrections to the Newtonian
potential.
\end{abstract}

\section{INTRODUCTION}

In a recent paper \cite{newton}
we have computed the corrections to the Newtonian potential
due to a quantum massive scalar field coupled to the metric in a
$R+ R^2$-theory
of gravitation. This
computation was
carried out by means of a non-local approximation to the Effective Action
(EA) \cite{vilk1,vilk2},
from which the effective
gravitational equations of motion were deduced. Expanding in
powers of $- \frac{m^2}{\Box}$, these equations read
\begin{eqnarray}
& &\left[
    \alpha_0 -
    \frac{1}{64 \pi^2} \left(
		  (\xi-\frac{1}{6})^2 - \frac{1}{90}
	       \right) \ln ({- \frac{\Box}{\mu^2}})
\right] H_{\mu\nu}^{(1)} +
\left[
    \beta_0 -
    \frac{1}{1920 \pi^2} \ln(-\frac{\Box}{\mu^2})
\right] H_{\mu\nu}^{(2)} + \nonumber\\
& &\left[
   - {1\over 8\pi G}+{m^2\over 16\pi^2}(\xi -{1\over 6})(-1
+ \ln {m^2\over \mu^2})
\right]
\left(R_{\mu\nu}-{1\over 2}Rg_{\mu\nu}\right) - \nonumber\\
& & \frac{m^2}{384 \pi^2}
    \ln(-\frac{\Box}{m^2}) \frac{1}{\Box}
    \left[
    (1-12 \xi^2) H_{\mu\nu}^{(1)}  - 2 H_{\mu\nu}^{(2)} \right]
    = O({\cal R}^2)
\label{eq:ec4d}
\end{eqnarray}
where $m$ is the mass of the scalar field, $\xi$ is the coupling
to the scalar curvature and
\begin{eqnarray}
H_{\mu\nu}^{(1)}&=& 4\nabla_{\mu}\nabla_{\nu}R - 4g_{\mu\nu}\Box R
+O({\cal R}^2)\nonumber\\
H_{\mu\nu}^{(2)}&=& 2\nabla_{\mu}\nabla_{\nu}R - g_{\mu\nu}\Box R
		    -2\Box R_{\mu\nu} +O({\cal R}^2)
\end{eqnarray}

The gravitational constants $\alpha_0 ,\beta_0$ and $G$
depend on the scale $\mu$ according with
the Renormalization Group
Equations (RGEs)
\cite{BD}
\begin{eqnarray}
\mu{d\alpha_0 \over d\mu}&=&
 -\frac{1}{32 \pi^2}
 \left[
 \left( \xi - \frac{1}{6}  \right)^2 -\frac{1}{90}
 \right] \label{eq:alfa} \\
\mu{d\beta_0 \over d\mu}&=&
 -\frac{1}{960 \pi^2} \label{eq:beta} \\
\mu{dG\over d\mu} & = &  \frac{G^2 m^2}{\pi} \left(\xi -  \frac{1}{6}  \right)
\label{eq:ge}
\end{eqnarray}
These are basically given by the Schwinger-DeWitt (SDW) coefficients
and can be obtained by imposing Eqn.(\ref{eq:ec4d})
to be independent of the renormalization scale $\mu$.
Comparing the RGEs with the effective Eqn.(\ref{eq:ec4d})
one readily notes that, while the corrections
proportional to $\ln(-\frac{\Box}{\mu^2})$ can be interpreted as
non-local modifications to
$\alpha_0$ and $\beta_0$,
this is not the case for the Newton constant.
Indeed, because of the identity
\begin{equation}
R_{\mu\nu} - \frac{1}{2} R g_{\mu\nu} = \frac{1}{4 \Box}
( H_{\mu\nu}^{(1)} - 2 H_{\mu\nu}^{(2)} ) + O({\cal R}^2)
\label{eq:nn}
\end{equation}
the non-analytic corrections proportional to $-\frac{m^2}{\Box}
\ln(-\frac{\Box}{m^2})$ can be
interpreted as modifying $G$ only for $\xi = 0$. This has also
been pointed out in Ref.\cite{PT}.

This discrepancy can also be seen at the level of the Newtonian
potential, which has $\frac{\log r}{r}$ and $r^{-3}$ quantum corrections
\cite{newton}.
The  $r^{-3}$ corrections come from the $\ln({-\Box\over \mu^2})$
terms in the effective equations and survive in the massless
limit (similar corrections due to the graviton
sector of the theory have been found in \cite{donoghue}).
The  $\frac{\log r}{r}$ corrections come from
the term proportional to $-\frac{m^2}{\Box}\ln(-\frac{\Box}{m^2})$.
In principle,  one could `derive' these
logarithmic corrections from the RGE (5), replacing in the classical
potential $V_{cl}(r)$ the Newtonian
constant by its running counterpart and identifying
$\mu \leftrightarrow r^{-1}$.
The resulting `Wilsonian' potential $V(r)=-G(\mu=r^{-1})/r$
coincides with the one obtained in Ref.\cite{newton} only
for minimal and conformal coupling.
\footnote{
The coincidence at  $\xi = 1/6$ takes place only after
tracing the equations of motion.}

The aim of this work is to elucidate the
origin of the discrepancy between the scaling behaviour of the
Newton constant deduced from the effective equations
of motion and that obtained
through the RGEs.
To this end we will show that there is a qualitative difference
between the non-local corrections proportional to
$\ln(-\Box)$ and those proportional to
$-\frac{m^2}{\Box}\ln(-\Box)$. While the former are linked to the
{\it early-time} behaviour of the heat kernel\cite{Gospel}
(and consequently are determined by the $\hat{a}_2$ SDW coefficient),
the latter depend on the {\it late-time} behaviour and
produce the above-mentioned discrepancy.
We will prove this claim in Section II, where
we will also extend the
four-dimensional results to arbitrary dimensions.
In Section III we will analyze the same problem for spinor fields.

We emphasize that throughout this paper we will consider
quantum matter fields on a classical gravitational background.
This
will be enough for our main discussion, since at this
{\it semiclassical} level
we already have running coupling constants and
quantum corrections to the field equations
and Newtonian potential. Therefore we can compare
both answers and look for the reason of the discrepancy.

In order to go beyond the semiclassical theory,
there are two alternatives.
If the $R+R^2$-theory is considered as an effective,
low-energy field theory \cite{simon,mazzi},
the inclusion of the graviton
sector
can be done along the lines of Ref.\cite{donoghue},
and we expect additional $r^{-3}$ corrections
to the Newtonian potential.
On the other hand, if the $R+R^2$-theory is considered
as a complete and renormalizable theory of gravity,
 due
to asymptotic freedom \cite{avra}, the graviton sector
could produce
an important increase of $G$ with distance \cite{alamos}.
However, in this case the $R+R^2$-theory is non-unitary,
and no definite conclusions can be drawn.
This point is beyond the scope of this paper.

\section {SCALING FOR SCALAR FIELDS}

Let us consider the evaluation of the one-loop contribution of a
massive quantum scalar field to the gravitational EA
\begin{equation}
\Gamma = \frac{1}{2} {\rm ln \, det}
( -\Box + m^2 + \xi R )
\end{equation}
The task of
evaluating this functional determinant on an arbitrary background is quite
complicated and approximation methods are compelling. Using the early-time
expansion of the heat kernel, the EA in $2 w$ dimensions
reads \cite{vilk1,vilk2,Gospel}
\begin{equation}
\Gamma = - \frac{1}{2} \lim_{L^2 \rightarrow \infty}
		 \frac{1}{(4 \pi)^{w}} \int_{1/L^2}^{\infty}
		 \frac{ds}{s^{w+1}} \exp(-s m^2) \sum_{n=0}^{\infty}
		 s^{n} \int d^{\scriptscriptstyle{2 w}}x \sqrt{g} \,
		\hat{a}_{n}(x)
\label{eq:kernel}
\end{equation}
where the ultraviolet divergence is regularized by the introduction of a
positive lower limit in the proper-time integral. Here all the functions
$\hat{a}_{n}(x)$ are the coincident limit of the SDW
coefficients.

As suggested by Vilkovisky
\cite{Gospel} , when the background fields are weak but rapidly varying,
one can obtain a non-local expansion of the EA
by summing all terms with a given power of the curvature
and any number of derivatives in the SDW series. The result
is well-behaved in
the massless limit and can be written as
\begin{equation}
\Gamma  =
- \frac{1}{2} \frac{1}{(4 \pi)^{w}} \int d^{\scriptscriptstyle{2 w}}x \sqrt{g}
\lim_{L^2 \rightarrow \infty}
\left(
      h_{0} + h_{1} (\frac{1}{6} - \xi) R +
	       R F_1(\Box) R +  R_{\mu\nu} F_2(\Box) R_{\mu\nu}
+  O({\cal R}^3)\right )
\label{eq:gamma2}
\end{equation}
where $h_{n} = \int_{1/L^2}^{\infty} ds s^{n-w-1} e^{-s m^2}$,
$F_i(\Box) = \int_{1/L^2}^{\infty} ds \frac{e^{-s m^2}}{s^{w-1}}
f_{i}(-s \Box)$
and the form factors $f_{i}$ are functions to be defined afterwards.

Up to here no assumptions about the mass $m$ have been made. In the large mass
limit, $m^2 {\cal R}\gg\nabla\nabla {\cal R}$, the SDW expansion
is recovered, while in the opposite one, the form factors can be expanded in
powers of $z \equiv - \frac{m^2}{\Box}$. We shall be working in the latter
limit. We have to evaluate the integral
\begin{equation}
I_{w} \stackrel{\rm def}{=}  \lim_{L^2 \rightarrow \infty}
\int_{1/L^2}^{\infty}
ds \frac{e^{- m^2 s}}{s^{w-1}} \sigma(-s \Box)
\end{equation}
where $\sigma$ denotes generically the $f_{i}$'s.
In order to study the
behaviour of $I_{w}$ in terms of the small quantity $z$, we split up the
integral into two terms
\begin{eqnarray}
I_{w} & = &  \lim_{L^2 \rightarrow \infty}
(A_{w} + B_{w}) \nonumber \\
A_{w} & = &
(-\Box)^{w-2} \int_{-\Box/L^2}^{C} \frac{d\eta}{\eta^{w-1}}
e^{-\eta z}\sigma(\eta)
\nonumber \\
B_{w} & = &
(-\Box)^{w-2} \int_{C}^{\infty} \frac{d\eta}{\eta^{w-1}}
e^{-\eta z} \sigma(\eta)
\end{eqnarray}
where $C$ is chosen such that $z^{-1} \gg C \gg 1$. Let us analyze the two
integrals separately.

For the $A_{w}$ integral, one can use the Taylor expansion of the form factor,
namely
$\sigma(\eta) = \sum_{n=2}^{\infty} \sigma_{n} \eta^{n-2}$.
The constants $\sigma_{n}$ can be read from the corresponding SDW
coefficient $\hat{a}_{n}$, as follows from
Eqns(\ref{eq:kernel},\ref{eq:gamma2}). The $n \ge w+1$ terms have a finite
$L^2 \rightarrow \infty$ limit that gives a $\Box$-dependent contribution
that is analytic in the variable $z$, while the $2 \le n \le w$ terms
are UV divergent. Expanding the exponential in $A_{w}$ in powers of the
small quantity $\eta z$ we obtain its final expression
\begin{eqnarray}
A_{w} & = & - (-\Box)^{w-2} Log(- \frac{\Box}{L^2})
\sum_{n=2}^{w}
\frac{\sigma_n}{(w-n)!} (-\frac{m^2}{\Box})^{w-n} + \nonumber \\
& & (-\Box)^{w-2} {\sum_{n=2}^{w} \sum_{k=0}^{w-n-1}}
\frac{\sigma_n}{(w-n-k) k!} (\frac{m^2}{\Box})^2 (-\frac{L^2}{\Box})^{w-n-k} +
\ldots
\label{eq:Aw}
\end{eqnarray}
where the dots denote finite terms, analytic in the small quantity
$-\frac{m^2}{\Box}$. Note that both the divergent and non-analytic parts
of $A_w$ are determined by the first $w$ SDW coefficients.
In order to renormalize the theory, the infinities have to be cancelled by
means of suitable counterterms in the classical lagrangian of the form
${\cal R} {\cal R} \, , \,
{\cal R} \Box {\cal R} \, , \, {\cal R} \Box^2 {\cal R} , \ldots ,
{\cal R} \Box^{w-2} {\cal R}$,
these being the only quadratic counterterms that can appear.
The UV divergences proportional to $\ln(L^2)$
that appear in both $A_w$ and the
$h_n$ integrals are absorbed in the bare constants, being renormalized by
terms of the form $\log(\frac{L^2}{\mu^2})$, where $\mu$ is an arbitrary
arbitrary scale parameter with units of mass. The fact that the EA must not
depend on this arbitrary parameter implies that the gravitational constants
scale with $\mu$, the scaling being given by the RGEs
(see Eqns(\ref{eq:alfa},\ref{eq:beta},\ref{eq:ge}) for the $w=2$ case).

As to the $B_{w}$ integral,
its leading behaviour in
powers of $- \frac{m^2}{\Box}$ is governed by the asymptotic expansion of the
form factor. Assuming that
$\sigma(\eta) = \frac{k}{\eta}$ as $\eta \rightarrow \infty$,
where $k$ is a numerical factor, the integral $B_{w}$ reads
\begin{equation}
B_w = k \frac{(-1)^w}{(w-1)!} (-\Box)^{w-2} (-\frac{m^2}{\Box})^{w-1}
\ln(-\frac{m^2}{\Box}) + \ldots
\label{eq:Bw}
\end{equation}
the dots being analytic terms.

Given the EA one can
derive the effective gravitational field equations. After a straightforward
calculation we find
\begin{eqnarray}
& & \left(
-\frac{1}{8 \pi G} + \frac{(-1)^w (m^2)^{w-1}}{(4 \pi)^w (w-1)!}
(\xi -\frac{1}{6}) \ln(\frac{m^2}{\mu^2})
\right)
( R_{\mu\nu} - \frac{1}{2} R g_{\mu\nu} )  +
\sum_{j=0}^{w-2}
\left[
\alpha_{j} \Box^{j} H_{\mu\nu}^{(1)} + \beta_{j} \Box^{j} H_{\mu\nu}^{(2)}
\right] = \nonumber \\
& & < T_{\mu\nu} > \stackrel{\rm def}{=}
-\frac{1}{2 (4 \pi)^w} [ F_{1}(\Box) H_{\mu\nu}^{(1)} +
					  F_{2}(\Box) H_{\mu\nu}^{(2)} ]
+ O({\cal R}^2)
\label{eq:ecmov}
\end{eqnarray}
In this equation the cosmological constant term has been omitted
and $\alpha_j$ and $\beta_j$ denote the
gravitational constants associated with the higher order terms in the
classical lagrangian.

In four-dimensional spacetime the basic integral $I_{w}$ can be calculated
using Eqns(\ref{eq:Aw},\ref{eq:Bw}). Up to analytic terms in
$-\frac{m^2}{\Box}$ it is given by
\begin{equation}
I_{\scriptscriptstyle w=2} = - \sigma_{2} Log(- \frac{\Box}{\mu^2}) -
k \frac{m^2}{\Box} Log(- \frac{m^2}{\Box})  +
O \left(-\frac{m^2}{\Box} \right)^2
\label{eq:Iw2}
\end{equation}
The corresponding stress tensor reads
\begin{eqnarray}
< T_{\mu\nu} > = \frac{1}{32 \pi^2}
\left(
\log(-\frac{\Box}{\mu^2}) [ \sigma_{2}^{(1)} H_{\mu\nu}^{(1)} +
			    \sigma_{2}^{(2)} H_{\mu\nu}^{(2)} ] +
\frac{m^2}{\Box} \log(-\frac{m^2}{\Box}) [ k^{(1)} H_{\mu\nu}^{(1)} +
					   k^{(2)} H_{\mu\nu}^{(2)} ]
\right)
\end{eqnarray}
the $\sigma_{2}^{(i)}$ and $k^{(i)}$ being the numerical constants in
Eqn(\ref{eq:Iw2}), respectively associated with the $R^2$ and
$R_{\mu\nu} R_{\mu\nu}$ terms in the EA.

The $m^2$-independent terms in $<T_{\mu\nu}>$ can be interpreted as being
quantum corrections to the gravitational constants $\alpha_0$ and $\beta_0$.
As was already mentioned, the numerical coefficients $\sigma_2^{(i)}$
associated with these corrections are basically given by the $\hat{a}_2$ SDW
coefficient (early-time behaviour of the heat kernel).
When the equations of motion are traced and solved,
these terms produce $r^{-3}$ quantum corrections to the Newtonian
potential [1].

In an analogous way, one would expect that the $m^2$-dependent terms in
$<T_{\mu\nu}>$, namely
\begin{equation}
\frac{ m^2 k^{(1)} }{32 \pi^2} \log(-\frac{m^2}{\Box}) \frac{1}{\Box}
( H_{\mu\nu}^{(1)} + \frac{ k^{(2)} }{ k^{(1)} } H_{\mu\nu}^{(2)} )
\label{eq:gcorr}
\end{equation}
could be expressed in a combination proportional to
$m^2 \log(-\frac{\Box}{m^2}) (R_{\mu\nu} - \frac{1}{2} R g_{\mu\nu})$, so
that they can be interpreted as a quantum correction to the Newton constant.
 From Eqn(\ref{eq:nn}) we see that
the aforementioned combination comes up only for
$k^{(2)}/k^{(1)} = -2$, a condition that is not always met. Also note that
the correction depends on the numerical coefficients $k^{(i)}$, which are
given by the asymptotic late-time behaviour of the heat kernel. The terms
in Eqn(\ref{eq:gcorr}) produce
a $\frac{\log r}{r}$ correction to the Newtonian potential [1].

The coefficients $\sigma_{n}$'s and $k$'s can be evaluated from the
form factors $f_{i}$'s. These are defined through the basic form factor
$f(\eta) = \int_{0}^{1} dt e^{-t(1-t) \eta}$ as follows \cite{vilk2,manitoba}
\begin{eqnarray}
f_{1}(\eta) & = & \frac{f(\eta)}{8}
\left[
\frac{1}{36} + \frac{1}{3 \eta} -
\frac{1}{\eta^2}
\right] - \frac{1}{16 \eta} + \frac{1}{8 \eta^2} +
(\xi - \frac{1}{6}) \left[
\frac{f(\eta)}{12} + \frac{ f(\eta)-1}{2 \eta} \right] +
\frac{1}{2} (\xi -\frac{1}{6})^2 f(\eta) \nonumber \\
f_{2}(\eta) & = & \frac{ f(\eta) - 1 + \eta/6 }{\eta^2}
\end{eqnarray}
 From here the relevant coefficients for the four-dimensional theory can be
calculated: $\sigma_2^{(i)} = f_i(0)$ and
$k^{(i)} = \lim_{\eta \rightarrow \infty} \eta f_i(\eta)$. Therefore we have
\begin{eqnarray}
\sigma_{2}^{(1)} = \frac{1}{2}
\left[ \left( \frac{1}{6} - \xi \right)^2 -\frac{1}{90} \right]
& ~~~~~ & \sigma_{2}^{(2)} = \frac{1}{60} \nonumber \\
k^{(1)} = \xi^2 - \frac{1}{12} & ~~~~~ &
k^{(2)} = \frac{1}{6}
\label{eq:coef4d}
\end{eqnarray}
It is straightforward to see that only for minimal coupling ($\xi=0$)
can the $m^2$-dependent part of $<T_{\mu\nu}>$ be interpreted as correcting
the Newton constant.

All this reasoning case be extended for
arbitrary values of $w$. All terms in the energy-momentum tensor that
depend on the $\sigma_n$'s can be interpreted as being quantum corrections
to the gravitational constants associated with the corresponding
${\cal R} \Box^{n-2} {\cal R} , (2 \leq n \leq w)$ terms in the classical
lagrangian. The numerical coefficients $\sigma_n$'s in these corrections
depend on the $\hat{a}_n$ SDW coefficient. On the contrary, the terms with
higher power
of the mass ($k$-dependent ones) involve the asymptotic behaviour
of the non-local form factors and can be viewed as correcting the Newton
constant only for $\xi=0$.

For example, in six-dimensional spacetime the integral $I_{w}$ can be
calculated using Eqns(\ref{eq:Aw},\ref{eq:Bw}) and is given by
\begin{equation}
I_{\scriptscriptstyle w=3} = \sigma_3 \Box Log(- \frac{\Box}{\mu^2}) +
\sigma_2 m^2 Log(-\frac{\Box}{\mu^2}) - k \frac{m^4}{2 \Box}
Log(- \frac{\Box}{m^2})
\label{eq:Iw3}
\end{equation}
For this theory the
coefficients $\sigma_2$ and  $k$ are the same as those of the four-dimensional
one, while the $\sigma_3$ coefficients are obtained from the term of the form
factors that is linear in $\eta$ and read
\begin{eqnarray}
\sigma_{3}^{(1)} = -\frac{1}{336} + \frac{\xi}{30} -
\frac{\xi^2}{12}
& ~~~~~~~~~ & \sigma_{3}^{(2)} = -\frac{1}{840}
\label{eq:coef6d}
\end{eqnarray}
In this case one obtains that the $m^0$ ($m^2$) terms
in $<T_{\mu\nu}>$
are interpreted as quantum corrections to the gravitational
coefficients $\alpha_0$, $\beta_0$ ($\alpha_1$, $\beta_1$) and depend on the
$\hat{a}_2$ ($\hat{a}_3$) SDW coefficient. As before, one can view the $m^4$
terms as a quantum correction to the Newton constant only for minimal
coupling.

Having evaluated the energy-momentum tensor, we shall make a brief
comment on the trace anomaly. As is well-known \cite{BD}, the
classical theory is
conformally invariant for $m=0$ and
$\xi = \frac{1}{4} \frac{2w-2}{2w-1}$. Due to quantum effects, a trace
anomaly in $< {T_{\mu}}^{\mu} >$ appears, which is local and proportional
to the $\hat{a}_w$ SDW coefficient. In our computation of the
energy-momentum tensor we have concentrated on the
non-local terms and we have absorbed the local ones into the renormalized
classical gravitational constants. Using the expressions for the
coefficients $\sigma_w^{(i)}$ evaluated at conformal coupling
(see Eqns(\ref{eq:coef4d},\ref{eq:coef6d}) for the $w=2$ and $w=3$ cases)
one can readily prove that
the trace of the non-local and mass-independent terms of the energy-momentum
tensor vanishes. Although the local terms are irrelevant for the main
point of this work, which is throughly developed in previous
paragraphs, their evaluation from the integral $A_w$ is straightforward.
At conformal coupling these terms give the correct trace anomaly, up to
the order we are working (the $O({\cal R}^2)$ contributions
to the trace anomaly have been recently computed from the non-local
effective action in Ref.\cite{trace})

\section{ SCALING FOR SPINOR FIELDS}

In this section we shall extend the reasoning to spinor fields
in four dimensions.
The one-loop contribution to EA of the free Dirac field
on a gravitational background is
\begin{eqnarray}
& \Gamma = - \frac{1}{2} {\rm Tr ln} \hat{K} &
\nonumber \\
\hat{K} \Psi  & =
(\gamma_{\mu} \nabla_{\mu} + m) (-\gamma_{\nu} \nabla_{\nu} + m) \Psi = &
(- \Box + m^2 + \frac{1}{4} R) \Psi
\end{eqnarray}
Therefore we have to evaluate the
trace of an operator similar to that associated with the scalar field
for $\xi = 1/4$ and trace over the spinor indexes.

We shall evaluate the EA following the method described in
the previous Section (see Eqn(\ref{eq:gamma2})).
The second order term in curvatures can be written as \cite{vilk1,vilk2}
\begin{equation}
\Gamma^{(2)} =  \frac{1}{32 \pi^2} \int d^{4}x \sqrt{g}
\left[
4 R F_{1}(\Box) R +
4 R_{\mu\nu} F_{2}(\Box) R_{\mu\nu} +
{\rm Tr} ( {\cal R}_{\mu\nu} F_{3}(\Box) {\cal R}_{\mu\nu} )
\right]
\label{eq:easpinor}
\end{equation}
where ${\cal R}_{\mu\nu}  = [\nabla_{\mu},\nabla_{\nu}] =
\frac{1}{8} [\gamma_{\alpha}(x),\gamma_{\beta}(x)]_{-}
R_{\alpha\beta\mu\nu}(x) $ is the commutator of the covariant derivatives
\cite{christ}.
Here $F_{1}(\Box)$ and $F_{2}(\Box)$ are the scalar field-form
factor integrals evaluated at $\xi =1/4$. We have a new
contribution proportional to $F_{3}(\Box) =
\int_{1/L^2}^{\infty} ds \frac{e^{-s m^2}}{s^{w-1}}
{f(-s \Box)-1\over 2 s\Box}$,
due to the non-vanishing
commutator of the covariant derivatives.

Using the expression for ${\cal R}_{\mu\nu}$
and calculating the trace of the product
of four gamma matrices, the last term in Eqn(\ref{eq:easpinor})
can be written as
${\rm Tr } {\cal R}_{\mu\nu} F_{3}(\Box) {\cal R}_{\mu\nu} =
- \frac{1}{2} R_{\alpha\beta\mu\nu} F_{3}(\Box) R_{\alpha\beta\mu\nu}
$.
Finally, using integration by parts, the Bianchi identities
and the non-local expansion of the
Riemann tensor in terms
of the
Ricci tensor \cite{vilk2,manitoba}
\begin{equation}
R_{\alpha\beta\mu\nu} = \frac{1}{\Box} \{
\nabla_{\mu} \nabla_{\alpha} R_{\nu\beta} +
\nabla_{\nu} \nabla_{\beta} R_{\mu\alpha} -
\nabla_{\nu} \nabla_{\alpha} R_{\mu\beta} -
\nabla_{\mu} \nabla_{\beta} R_{\nu\alpha}  \} + O( {\cal R}^2 )\,\,\,  ,
\end{equation}
one can rewrite the last expression
through a kind of
generalized Gauss-Bonnet identity, namely
\begin{equation}
\int d^{4}x {\rm Tr } {\cal R}_{\mu\nu} F_{3}(\Box) {\cal R}_{\mu\nu} =
\int d^{4}x
\left[
\frac{1}{2} R F_{3}(\Box) R -
2 R_{\mu\nu} F_{3}(\Box) R_{\mu\nu} + O({\cal R}^3)
\right]
\end{equation}
In view of this identity, the stress tensor is basically the one for
the scalar
field, modified as follows
\begin{eqnarray}
< T_{\mu\nu} > = - \frac{1}{32 \pi^2}
&  & \left(
\log(-\frac{\Box}{\mu^2})
\left[
(4 \sigma_{2}^{(1)} + \frac{1}{2} \sigma_{2}^{(3)}) H_{\mu\nu}^{(1)} +
(4 \sigma_{2}^{(2)} - 2 \sigma_{2}^{(3)}) H_{\mu\nu}^{(2)}
\right] +
\right. \nonumber \\
& & \left.
\frac{m^2}{\Box} \log(-\frac{m^2}{\Box})
\left[
(4 k^{(1)} + \frac{1}{2} k^{(3)}) H_{\mu\nu}^{(1)} +
(4 k^{(2)} - 2 k^{(3)}) H_{\mu\nu}^{(2)}
\right]
\right)
\label{eq:dirac}
\end{eqnarray}
The new coefficients, associated to the form factor integral $F_{3}$,
are given by $\sigma_{2}^{(3)} = 1/12$
(early-time behaviour) and $k^{(3)} =1/2$ (late-time behaviour),
and the other
coefficients, written in Eqn(\ref{eq:coef4d}),
are evaluated at $\xi=1/4$. Therefore the $m^2$-dependent terms
in $<T_{\mu\nu}>$ can be seen as correcting the Newton constant since
$(4 k^{(2)} - 2 k^{(3)}) / (4 k^{(1)} + \frac {k^{(3)}}{2} ) = -2$.
The spinor field behaves, in this respect, as the minimally-coupled scalar
field.

Finally, after tracing and solving the equations of motion, the quantum
correction to the Newtonian potential reads
$\delta V(r)  = - \frac{G^2 M m^2}{3 \pi} \frac{\log\frac{r}
{r_{\scriptscriptstyle 0}}}{r} $
which coincides with the Wilsonian potential, obtained from the RGE for the
Newton constant $G(\mu)$.

\section{Acknowledgments}

F.D.M. would like to thank C. Fosco for useful discussions
on related matters, and IAEA and UNESCO for
hospitality at ICTP.
This research was supported by Universidad de Buenos Aires,
Consejo Nacional de Investigaciones Cient\'\i ficas y T\' ecnicas
and by Fundaci\' on Antorchas.

\end{document}